\documentstyle[12pt,epsfig]{article}

\newcommand{\lsim}{\mathrel{\lower4pt\hbox{$\sim$}}
\hskip-12.5pt\raise1.6pt\hbox{$<$}\;}
\newcommand{\gsim}{\mathrel{\lower4pt\hbox{$\sim$}}
\hskip-12.5pt\raise1.6pt\hbox{$>$}\;}

\newcommand{\bea}{\begin{eqnarray}}
\newcommand{\eea}{\end{eqnarray}}

\newcommand{\be}{\begin{equation}}
\newcommand{\ee}{\end{equation}}

\begin{document}
\begin{titlepage}
\begin{flushright}
CERN--TH/2000-070\\
LNF--00/005(P)\\
ULB--TH/99-18\\
hep-ph/0003052
\end{flushright}
\vspace*{1.6cm}

\begin{center}
{\Large\bf Large $N_c$, chiral approach to $M_{\eta^\prime}$ at finite temperature}\\
\vspace*{0.8cm}

R.~Escribano$^1$, F.S.~Ling$^2$, and M.H.G.~Tytgat$^3$\\
\vspace*{0.2cm}

{\footnotesize\it 
$^1$INFN-Laboratori Nazionali di Frascati, P.O.~Box 13, I-00044 Frascati, Italy\\
$^2$Service de Physique Th\'eorique, Universit\'e Libre de Bruxelles, 
CP 225, B-1050 Bruxelles, Belgium\\
$^3$CERN-Theory Division, CH-1211 Geneva 23, Switzerland}
\end{center}
\vspace*{1.0cm}

\begin{abstract}
We study the temperature dependence of the $\eta$ and $\eta^\prime$ meson masses 
within the framework of $U(3)_L \times U(3)_R$ chiral perturbation theory, 
up to next-to-leading order in a simultaneous expansion in momenta, quark masses 
and number of colours.
We find that both masses decrease at low temperatures, but only very slightly.
We analyze higher order corrections and argue that large $N_c$ suggests a 
discontinuous drop  of $M_{\eta^\prime}$ at the critical temperature of 
deconfinement $T_c$, consistent with a first order transition to a phase with 
approximate $U(1)_A$ symmetry.
\end{abstract}

\vspace*{1.6cm}
\begin{flushleft} CERN--TH/2000-070\\
March 2000
\end{flushleft}
\end{titlepage}

\section{Introduction}

The fate of the $U(1)_A$ symmetry of QCD at finite temperature is a fascinating 
problem~\cite{Gross:1981br,Pisarski:1984ms,Shuryak:1994ee} which could  also 
have interesting consequences for the ongoing heavy ion collisions program and, 
possibly, for cosmology~\cite{Kapusta:1996ww,huang,schafer,Kharzeev:1998kz,
Schaffner-Bielich:1999uj}.
Even a partial restoration of $U(1)_A$ symmetry in the vicinity of the critical 
temperature of chiral symmetry breaking ($T_c \sim 200$ MeV)  
could dramatically change the mass and mixing pattern of the lightest neutral 
mesons ($\pi^0$, $\eta$ and $\eta^\prime$), with signals including enhanced 
strangeness production or the more speculative possibility of forming parity 
violating disoriented $\eta^\prime$ condensates in heavy ion 
collisions~\cite{Kharzeev:1998kz,Morley:1985wr,Buckley:1999mv}. 

Our aim in the present paper is rather modest: 
we will study the shift of the mass of the $\eta$ and $\eta^\prime$ mesons at 
low temperatures, in a regime in which the hadronic gas is mostly composed of 
pions. 
We will work in the framework of $U(3)_L\times U(3)_R$ chiral perturbation theory 
($\chi PT$), in a simultaneous expansion in momenta, quark masses, number of 
colours $N_c$, and temperature $T$. 

Our motivation for doing this investigation was threefold. 
First, the predictions of $\chi PT$ in a pion thermal bath, although limited in 
scope to $T \lsim \mbox{\rm few}\ f_\pi$, are essentially model independent 
(see for instance, the review of Smilga~\cite{Smilga:1996cm} and references 
therein). 
Given the phenomenological success of the large $N_c$ expansion in vacuum, 
one might 
perhaps hope 
that the predictions 
of 
the present work are as robust.  
Next, we wanted to see to which extent the results derived in 
Ref.~\cite{Jalilian-Marian:1998mb} could be amended. 
As us, the authors have computed the shift of $M_\eta$ and $M_{\eta^\prime}$ at
low temperatures using the Di Vecchia-Veneziano-Witten effective lagrangian 
(DVW)~\cite{Witten:1980sp,DiVecchia:1980ve,Rosenzweig:1980ay}, but only to 
leading order $\Delta M^2_{\eta^\prime} \sim T^2$.
However, it was not clear to us whether the leading order DVW lagrangian was a 
good approximation for this problem. 
Although the parameters of the lagrangian can be fitted to the observed mass 
and mixing pattern of the $\eta$ and $\eta^\prime$ mesons to within 
$10 \%$~\cite{Georgi:1993jn,Peris:1994ga}, the decay rates predicted for
$\eta^\prime \rightarrow \eta \pi \pi$ are off the experimental values by a 
factor of about $40$. This issue, which is obviously relevant in order to 
determine the shift of $M_{\eta^\prime}$ in a pion bath, is however easily 
cured at next-to-leading order in the large $N_c$ 
expansion~\cite{Herrera-Siklody:1999ss,DiVecchia:1981sq}. 
As we will show, next-to-leading order corrections are also quite important
at finite temperature, but not to the point of dramatically changing the 
conclusion of Ref.~\cite{Jalilian-Marian:1998mb}: at low temperatures, 
$M_{\eta^\prime}$ stays essentially constant.
Finally, we wanted to see  what the large $N_c$ expansion could teach us 
about the fate of the $U(1)_A$ symmetry at finite temperature. 
At zero temperature, in the confined phase of QCD, large $N_c$ arguments 
predict that $M_{\eta^\prime}^2 \propto 1/N_c$. 
On the other hand, at very high temperatures, in the quark-gluon plasma phase 
of QCD ($T \gg \mu_{\rm hadr} \sim 200$ MeV), instanton calculus is reliable 
and predicts an effective restoration of $U(1)_A$ symmetry. 
Because of screening, instanton effects are suppressed at very large 
temperatures $\propto \exp(-8 \pi^2/g(T)^2)$.
At large $N_c$, the suppression is more important, 
as $1/g^2 \rightarrow N_c/ \lambda$ with fixed 't Hooft coupling 
$\lambda= g^2 N_c$, and the exponential tends to vanish 
$\exp(-N_c/\lambda) \rightarrow 0$ as $N_c$ increases. 
Because of asymptotic freedom, $\lambda$ growths at lower temperatures and 
the instanton argument breaksdown. 
However, for $N_c$ large enough, a natural assumption is that the exponential 
suppression  holds  all the way down to the critical temperature of 
deconfinement $T_c\sim \mu_{\rm hadr}$~\cite{Kharzeev:1998kz}. Although 
we have no proof of this statement, such a behaviour seems natural given the 
large release of entropy $\propto N_c^2$ at $T_c$ and is actually known to 
occur in  models in two dimensions~\cite{Affleck:1980gy}. 
With this assumption, $M_{\eta^\prime}$ can be taken as an order parameter 
for $U(1)_A$ symmetry restoration at $T_c$.
In Ref.~\cite{Kharzeev:1998kz}, some information on the behaviour of 
$M_{\eta^\prime}$ near $T_c$ could be extracted assuming that the 
{\em deconfining} phase transition could be of second order at large 
$N_c$~\cite{Pisarski:1997yh}. 
We will argue here that large $N_c$ favours a sharp drop of $M_{\eta^\prime}$ 
at $T_c$, consistent with first order transition to the phase with 
(approximate) $U(1)_A$ symmetry.

Our paper is organized as follows. In the next section, we briefly review  
$U(3)_L \times U(3)_R$ $\chi PT$, which extends the framework of the large 
$N_c$ DVW effective lagrangian beyond leading order. 
For definitiveness, we refer to the recent analysis of 
Herrera-Sikl\'ody {\em et al.}~\cite{Herrera-Siklody:1997pm}. 
We then discuss the implications of these corrections at low temperature, in 
presence of a pion thermal bath. 
Most formulas are relegated to the appendix. 
In the last section, we speculate on the effect of higher order corrections 
in the large $N_c$ expansion and draw the conclusions. 

\section{Sketch of $U(3)_L\times U(3)_R$ chiral perturbation theory}

At low energies and temperatures, the dynamics of QCD is governed by an 
approximate $SU(3)_L \times SU(3)_R$ chiral symmetry which is spontaneously 
broken to the diagonal $SU(3)$ in vacuum. 
If the mass of the up, down and strange quarks were vanishing, the symmetry
would become exact and there would be eight massless Goldstone bosons. 
Phenomenological lagrangians, which treat the mass of the quarks as small 
perturbations, provide a powerful framework, known as Chiral Perturbation 
Theory ($\chi PT$), to study the properties of the light $\pi$, $K$, and 
$\eta$ mesons~\cite{cpt,Gasser:1985gg}. 
The $\eta^\prime$ meson doesn't {\em a priori} fit in this frame. 
It is substantially heavier than the other eight light mesons, and, in vacuum,
would stay so even in the chiral limit of zero quark masses, because it    
receives most of its mass from the $U(1)_A$ anomaly through non-perturbative 
instanton-like effects~\cite{gerard}. 
The effect of the axial anomaly can however be conveniently turned off by 
going to the limit of large number of colours 
$N_c$~\cite{gnc,Witten:1979vv,Veneziano:1979ec}.  
At infinite $N_c$ and in the chiral limit, the global symmetry becomes 
$U(3)_L \times U(3)_R$, spontaneously broken in vacuum to $U(3)$, with nine 
massless Goldstone bosons. 
Like chiral symmetry breaking effects by finite quark masses, $1/N_c$ 
suppressed contributions can be systematically introduced as perturbations 
in an effective lagrangian, an approach which has been quite 
fruitful~\cite{Witten:1980sp,DiVecchia:1980ve,DiVecchia:1981sq,Leutwyler:1996sa}.
A systematic analysis of next-to-leading corrections, including ${\cal O}(p^4)$ 
operators, has been initiated in the 
recent~\cite{Herrera-Siklody:1999ss,Herrera-Siklody:1997pm,Herrera-Siklody:1997kd}. 
We refer to these latter works for more details and follow their conventions 
for ease of reference. 
We will work in  Euclidean spacetime with metric 
$g_{\mu\nu} \equiv \delta_{\mu\nu}$ and use the imaginary time formalism to 
compute the thermal corrections. 

\subsection{Leading order}

The leading order effective lagrangian is 
well-known~\cite{Witten:1980sp,DiVecchia:1980ve,Rosenzweig:1980ay}. 
In the notation of~\cite{Herrera-Siklody:1997pm} it is written as
\bea
\label{lol}
{\cal L}_{\rm LO} &=& \frac{f^2}{4}\Big( v_{02} X^2 + 
\langle \partial_{\mu}U^{\dagger}\partial_{\mu} U \rangle -
\langle U^{\dagger} \chi+ \chi^{\dagger} U \rangle \Big)\ ,
\eea
where $U$ is the $U(3)$ matrix
\begin{equation}
U  \equiv e^{i \sqrt{2} \Phi/f}\ ,
\end{equation}
with $\Phi$ the pseudoscalar meson matrix
\begin{equation}
\Phi = \left(
\begin{array}{ccc}
{\pi^0 \over \sqrt 2}+{\eta_8 \over \sqrt 6}+{\eta_0 \over \sqrt 3} & \pi^+ & K^+ \\[1ex]
\pi^- & -{\pi^0 \over \sqrt 2}+{\eta_8 \over \sqrt 6}+{\eta_0 \over \sqrt 3} & K^0 \\[1ex]
K^- & \bar K^0 & - {2\eta_8 \over \sqrt 6} + {\eta_0 \over \sqrt 3} 
\end{array}
\right)\ ,
\end{equation}
and $f=f_\pi=92.4$ MeV at leading order.
The mass matrix is
\be
\chi =  2\,B\,\mbox{diag}(m_u,m_d,m_s)\ ,
\ee
but we shall neglect isospin breaking effects ($m_u=m_d\equiv m$). 
The constant $B$ is related to the value of the $\langle\bar q q\rangle$ condensate,
$M_\pi^2 = 2 m B \equiv - 2m \langle \bar q q\rangle/f_\pi^2$ at leading order.
The combination 
\begin{equation}
X(x) \equiv \langle \log U(x)\rangle + i \theta_{\rm QCD} = 
i \, {\sqrt{6}\over f} \eta_0 + i \theta_{\rm QCD}\ ,
\end{equation}
is invariant under $U(3)_L \times U(3)_R$ transformations,
$\langle \log U \rangle \rightarrow \langle \log (g_R U g_L^\dagger) \rangle +
2 i \langle \alpha \rangle$ and $\theta_{\rm QCD}\rightarrow 
\theta_{\rm QCD} - 2 \langle \alpha \rangle$. 
Because of this, any arbitrary function of $X$  can {\em a priori} enter in the 
construction of the effective lagrangian, with thus little predictive power. 
This is  where the large $N_c$ expansion comes to the rescue by limiting the number 
of operators that can contribute at each level of approximation.  
In the chiral limit ($m, m_s \rightarrow 0$), Eq.~(\ref{lol}) gives
\begin{equation}
M_{\eta^\prime}^2 =  - 3 v_{02}\ ,
\end{equation}
which is the celebrated Veneziano-Witten relation for three massless 
flavours \cite{Witten:1979vv,Veneziano:1979ec}, with
$v_{02} \equiv - 2 \tau/f^2 \sim {1/ N_c}$,
where $\tau$ is the topological susceptibility of pure Yang-Mills theory. 
The rationale of $U(3)_L \times U(3)_R$ $\chi PT$ is to count powers of 
$p^2$, $m_q$, and $1/N_c$ on the same level
${\cal O}(\delta)$~\cite{Herrera-Siklody:1997pm,Leutwyler:1996sa}:
\begin{equation}
{\cal O}(\delta) \sim p^2 \sim m_q \sim 1/N_c\ .
\end{equation}
According to this counting rule, the leading order lagrangian~(\ref{lol}) is 
${\cal O}(\delta^0)$ because $f^2 \sim {\cal O}(N_c)$~\footnote{Note that the 
field expansion of $U$ brings further powers of $1/f \sim 1/\sqrt N_c$. 
The ${\cal O}(\delta)$ counting is to be understood to hold at the operator 
level.}.

At leading order, there are four unknown parameters in the lagrangian:
$f$, $v_{02}$, and the combinations $m B$ and $m_s B$ (or $x\equiv m_s/m-1$). 
On the other hand, we have at our disposal seven observables: $f_\pi$, $f_K$, 
the four masses of the light mesons, and the $\eta$--$\eta^\prime$ mixing angle $\theta$.  
Using $M_{\eta^\prime}$ as input and the formul\ae~given for reference in the appendix, 
one obtains
\begin{equation}
\begin{array}{lcl}
f^2  & = & f_\pi^2 = f_K^2\ ,\\[1ex]
2 m B & = & M_\pi^2\ ,\\[1ex]
x & \simeq & 24.1\ ,\\[1ex]
v_{02} & \simeq & - 0.22\ {\rm GeV}^2\ ,
\end{array}
\end{equation}
which predict that $\theta \simeq - 20^\circ$ and
\begin{equation}
M_\eta \simeq  494.4\ {\rm MeV}\ .
\end{equation}
Remarkably, the latter number is only 10\% off the experimental value 
$M_\eta = 547.3$ MeV. It is however known that adjusting the  parameters cannot improve
the prediction because the ratio $M_\eta^2/M_{\eta^\prime}^2$ has an upper 
bound\footnote{Assuming $M_\pi= 0$ to simplify, Eq.~(\ref{lol}) gives
\begin{equation}
{M_\eta^2 \over M_{\eta^\prime}^2} = 
{3 - y - \sqrt{9 + 2 y + y^2}\over 3 - y + \sqrt{9 + 2 y + y^2}}\ ,
\end{equation}
where $y \equiv 9 v_{02}/2(M_K^2 -M_\pi^2)$. This ratio reaches a maximum at 
$y = - 3$ (note that $v_{02} < 0$) corresponding to
\begin{equation}
{M_\eta \over M_{\eta^\prime}} <  0.518\ ,
\end{equation}
to be compared with the measured ratio $M_\eta / M_{\eta^\prime} \simeq 0.571$. Taking 
into account $M_\pi \not= 0$ improves things, but not enough.}~\cite{Georgi:1993jn}.
One has to take into account next-to-leading order corrections to reach
agreement~\cite{Peris:1994ga}.  

At leading order, the only coupling between $\eta^\prime$ and the pions is from the 
quark mass term in the lagrangian~(\ref{lol}) and is thus chirally suppressed. 
The amplitude for $\eta^\prime \rightarrow \eta \pi \pi$ is then
\begin{equation}
\label{adler}
{\cal A} = \frac{M^2_{\pi}}{6 f_{\pi}}
\Big( 2 \sqrt 2 \cos( 2 \theta) - \sin (2 \theta)\Big)\ .
\end{equation}
The corresponding  decay rates
\begin{equation}
\begin{array}{ccl}
\Gamma (\eta'\rightarrow \eta \pi^0 \pi^0) &=&  1.0 \ {\rm keV} \ , \\[1ex]
\Gamma (\eta'\rightarrow \eta \pi^+ \pi^-) &=&  1.9 \ {\rm keV} 
\approx  2 \times \Gamma (\eta'\rightarrow \eta \pi^0 \pi^0) \ ,
\end{array}
\end{equation}
are however much smaller than the experimental ones,
\begin{equation}
\begin{array}{ccc}
\Gamma_{\rm exp} (\eta'\rightarrow \eta \pi^0 \pi^0) &=&  
42.0 \pm 4.2 \ {\rm keV} \ , \\[1ex]
\Gamma_{\rm exp} (\eta'\rightarrow \eta \pi^+ \pi^-) &=&  
88.9 \pm 7.6 \ {\rm keV} \ .
\end{array}
\end{equation}
We will not speculate on the reasons for this well-known discrepancy 
(see Refs.~\cite{DiVecchia:1981sq,Fariborz:1999gr} for a more recent discussion), 
but simply note that within the present framework, this issue can also be resolved at 
next-to-leading 
order~\cite{Herrera-Siklody:1999ss,DiVecchia:1981sq}\footnote{Note 
that the amplitude (\ref{adler}) is constant and vanishes in the limit $m_u=m_d=0$,
for any $m_s$. However, general arguments~\cite{DiVecchia:1981sq} 
(and a fit to experimental data) indicate that for $m_u=m_d=0$, the amplitude should
behave like ${\cal A} = \mbox{\rm const} \times p_\pi^{(1)}\cdot p_\pi^{(2)}$, where
$p_\pi^{(1,2)}$ are the momenta of the outgoing pions, and where the constant is 
vanishing as the strange quark mass goes to zero. 
As shown in Ref.~\cite{DiVecchia:1981sq}, this behaviour can be easily accommodated
by introducing higher-order terms, an approach that is systematized by the 
$\delta$ expansion~\cite{Herrera-Siklody:1999ss}. 
The smallness of the leading order contribution is then considered as a mere
accident, related to the smallness of the ratio $M_\pi^2/M_K^2$.}.

\subsection{Next-to-leading order}

In our case, at next-to-leading order, ${\cal O}(\delta)$, only a few more terms 
can be added to the lagrangian (\ref{lol})~\cite{Herrera-Siklody:1997pm}:
\begin{equation}
\label{lag}
\begin{array}{rcl}
{\cal L}_{NLO}&=& {\cal L}_{LO} \\[1ex] 
&+& \frac{f^2}{4} \
\Big( -v_{31}\, X \langle U^{\dagger} \chi- \chi^{\dagger} U \rangle + 
v_{40}\, \langle U^{\dagger} \partial_{\mu}U \rangle 
\langle U^{\dagger} \partial_{\mu}U \rangle \\[1ex] 
&+& i\,v_{50}\ \langle U^{\dagger} \partial_{\mu}U \rangle 
\partial_{\mu}\theta_{\rm QCD} +
v_{60}\, \partial_{\mu}\theta_{\rm QCD} 
\partial_{\mu}\theta_{\rm QCD} \Big)  \\[1ex] 
&-& M_0 \, O_0 - M_3 \, O_3 + L_5 \, O_5 - L_8 \, O_8\ ,
\end{array}
\end{equation}
where the $O_{0,3,5,8}$ are ${\cal O}(p^4)$ operators whose coupling constants
are ${\cal O}(N_c)$:
\begin{equation}
\begin{array}{rcl}
O_0 &=& \langle \partial_\mu U \partial_\nu U^\dagger 
\partial_\mu U \partial_\nu U^\dagger \rangle \ , \\[1ex]
O_3 &=& \langle \partial_\mu U^\dagger \partial_\mu U 
\partial_\nu U^\dagger \partial_\nu U \rangle \ , \\[1ex]
O_5 &=& \langle \partial_\mu U^\dagger \partial_\mu U 
(U^\dagger \chi + \chi^\dagger U) \rangle \ , \\[1ex]
O_8 &=& \langle \chi^\dagger U \chi^\dagger U + 
U^\dagger \chi U^\dagger \chi \rangle \ .
\end{array}
\end{equation}

The couplings $v_{40}$, $v_{50}$, and $v_{60}$ are not independent and either one 
of them can be set to zero by  an appropriate change of variables, 
$\eta_0/f \rightarrow \eta_0/f + \kappa\,\theta_{\rm QCD}$.
We shall choose $v_{40} = 0$. Moreover, $v_{50}$ and $v_{60}$ will not appear in 
our calculations and can be discarded. At ${\cal O}(\delta)$, the only coupling
related to the breaking of $U(1)_A$ symmetry is thus $v_{31} \sim {\cal O}(1/N_c)$.
Note that the corresponding operator is also chirally suppressed $\propto m_q$.

At next-to-leading order, seven unknown parameters enter in the definition of the 
meson mass matrix: $f$, $v_{02}$, $v_{31}$, $L_{5,8}$, together with the quark 
masses $m$ and $m_s$  (see appendix for details). 
These can be expressed in terms of seven independent observables: $f_\pi$, $f_K$, 
$M_\pi$, $M_K$, $M_\eta$, $M_\eta^\prime$, and the $\eta$--$\eta^\prime$ mixing 
angle $\theta$~\cite{Herrera-Siklody:1997kd}. 
At this level, large $N_c$ $\chi PT$ is thus  not predictive. The strategy adopted
in Ref.~\cite{Herrera-Siklody:1997kd} was to impose that ${\cal O}(\delta)$ 
corrections are not too large so that the large $N_c$ expansion makes sense.
For  mixing angle in 
the range $20^\circ < \theta < 24^\circ$, the fit 
gives~\cite{Herrera-Siklody:1997kd}
\begin{equation}
\begin{array}{rcl}
0.980  \leq & {2 m B/ M^2_\pi} & \leq  0.988 \ , \\[1ex]
18.3 \leq & x & \leq  20.9 \ , \\[1ex]
0.214\ {\rm GeV}^2 \leq & \vert v_{02}\vert & \leq  0.239\ {\rm GeV}^2 \ , \\[1ex]
1.35 \ 10^{-3} \leq & L_8 &\leq  1.57 \ 10^{-3} \ , \\[1ex]
-0.164 \leq & v_{31} & \leq  -0.161 \ .
\end{array}
\end{equation}
together with $f = 90.8$ MeV and $L_5 = 2.0 \ 10^{-3}$ which are fixed by $f_\pi$ 
and $f_K$. 
Note that if $v_{02}$ does only change by about 10\%, the shift in $m_s$ is quite 
large, $\sim$ 20--25.

Because they have four derivatives, the operators $O_0$ and $O_3$ do not contribute 
to the meson mass matrix in vacuum. However, they give the dominant contributions  
to the decay $\eta^\prime \rightarrow \eta \pi \pi$~\cite{Herrera-Siklody:1999ss}.
This is essentially because the extra derivatives introduce large amplification 
factors, $\propto (M_{\eta^\prime}/M_\pi)^2$, with respect to the leading order 
amplitude\footnote{This may actually casts some doubts on the validity of $\chi PT$ 
for such processes as one could expect higher-order effects to give non-negligible
contributions to the decay $\eta^\prime \rightarrow \eta\pi\pi$. 
One may nevertheless hope that the large $N_c$ expansion is still reliable and that 
these corrections are $1/N_c$ suppressed. Whether this is true is unfortunately hard 
to check as we would evidently have too few hadronic data to completely fit the
parameters of the effective lagrangian at higher orders in the $\delta$ expansion. 
Of course, this is precisely why the large $N_c$ expansion is invoked in the first
place.}.
The observed decay rates are well reproduced with 
\begin{equation}
\begin{array}{rcl}
M_0 & \simeq & 1.2 \ 10^{-3} \ , \\[1ex]
M_3 & \sim & - 0.4 \ 10^{-3} \ ,
\end{array}
\end{equation}
values which can be independently inferred from the known $L_1$, $L_2$, and $L_3$ of 
$SU(3)\times SU(3)$ $\chi PT$ (in the nomenclature of Gasser and 
Leutwyler~\cite{Gasser:1985gg})\footnote{According to 
Ref.~\cite{Herrera-Siklody:1999ss}, 
$M_0 = {2 \over 3} (L_1 + L_2) + {\cal O}(N_c^0)$ and $M_3 = L_3 + 2 M_0$.}.

Thus all the parameters of the next-to-leading order effective lagrangian are fixed 
by low-energy phenomenology. 

\section{$M_\eta^\prime$ in a pion thermal bath}

In Ref.~\cite{Jalilian-Marian:1998mb}, the leading order lagrangian~(\ref{lol})
has been used to study the shift of $M_\eta$ and $M_{\eta^\prime}$ at
one-loop in a pion thermal bath. The effect they found is very tiny, as at 
$T \sim 200$ MeV the relative mass shifts are only ${\cal O}(0.1 \%)$. 
The reason for this is easy to understand. 
The $\eta$ and $\eta^\prime$ mesons receive most of their mass from the topological 
susceptibility term $\propto v_{02}$ and/or from the strange quark mass, while the
pion thermal corrections only modifies the tiny contribution from the pion mass term 
$\propto M_\pi^2$.  
Thermal  kaons could give a larger effect, $\propto M_K^2$, but the density of these 
is exponentially suppressed at low temperatures, $\propto \exp(-M_K/T)$. 
One might wonder whether next-to-leading order corrections could directly affect 
the contribution of the leading order $U(1)_A$ breaking term $v_{02}$. 
As we have seen in the previous section, five extra operators appear at
next-to-leading order in the large $N_c$ expansion and, of these, only the one with 
coupling $v_{31}$ is related to $U(1)_A$ symmetry breaking.
Unfortunately, this term is also chirally suppressed, $\propto m$, and its 
contribution is only ${\cal O}(v_{31} M_\pi^2 T^2/f_\pi^2)$. At temperatures of 
interest, this is small compared to $v_{02}$, but of the same magnitude as the 
leading order thermal correction. The other four operators will also contribute, 
but in a less interesting way, as they are invariant under $U(1)_A$. Furthermore, 
their effects are also $\propto M_\pi^2$.  

We have computed the shift of the mass of $\eta$ and $\eta^\prime$ at one-loop, 
at next-to-leading order in the expansion in $\delta$. We have {\em not} taken into 
account two-loop corrections from the leading order lagrangian. Although it is not 
clear whether this is legitimate numerically speaking, neglecting these is however 
consistent with the rules of large $N_c$ chiral perturbation theory. Indeed, the 
natural extension of $\delta$ power-counting to finite temperature is
\begin{equation}
{\cal O}(\delta) \sim p^2 \sim m_q \sim 1/N_c \sim T^2\ .
\end{equation}
At leading order, $M^2_{\eta^\prime} = {\cal O}(\delta) \sim 1/N_c$ and the one-loop 
thermal correction is $\propto M_\pi^2 T^2/f^2 \sim \delta^3$. At two-loop, using 
the leading order lagrangian, the shift is $\propto M^2_\pi T^4/f^4 \sim \delta^5$, 
while at one-loop using the next-to-leading order lagrangian, the shift is typically
$v_{31} M_\pi^2 T^2/f^2 \sim \delta^4$ (using $v_{31} \sim 1/N_c$) and thus dominant. 
Consistency thus requires to neglect the two-loop contributions. This greatly 
simplifies the calculations which are a bit cumbersome, but otherwise straightforward.
\begin{figure}[t]
\centerline{\epsfig{file=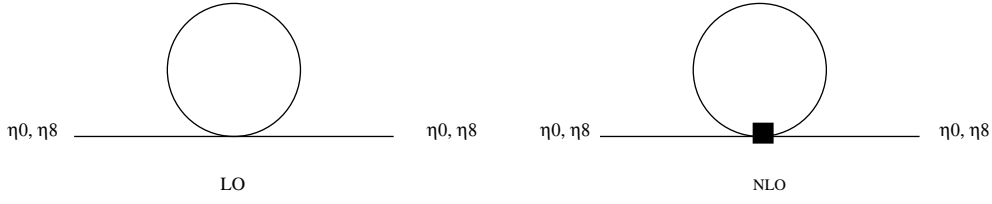,height=2.5cm}}
\caption{One-loop pion corrections to $M_\eta$ and $M_{\eta^\prime}$ at 
low temperature. The black box represent insertions of next-to-leading 
order operators.}
\label{fig1}
\end{figure}

The relevant diagrams are those of Fig.~\ref{fig1}, where the loops contain only pions.
At next-to-leading order there are two related thermal loops:
\begin{equation}
I_1(T) = T \sum_{n=-\infty}^\infty 
\int {d^3 \vec k\over (2\pi)^3}\, {1\over K^2 + M_\pi^2}\ ,
\end{equation}
with $K^2=k_0^2 + \vec k^2$, where $k_0= 2 \pi n T$, with $n$ integer, 
are the Matsubara frequencies and
\begin{equation}
I_2(T) =  T \sum_{n=-\infty}^\infty 
\int {d^3 \vec k\over (2\pi)^3}\, {K^2 \over K^2 + M_\pi^2}\ ,
\end{equation}
\begin{figure}[t]
\centerline{\epsfig{file=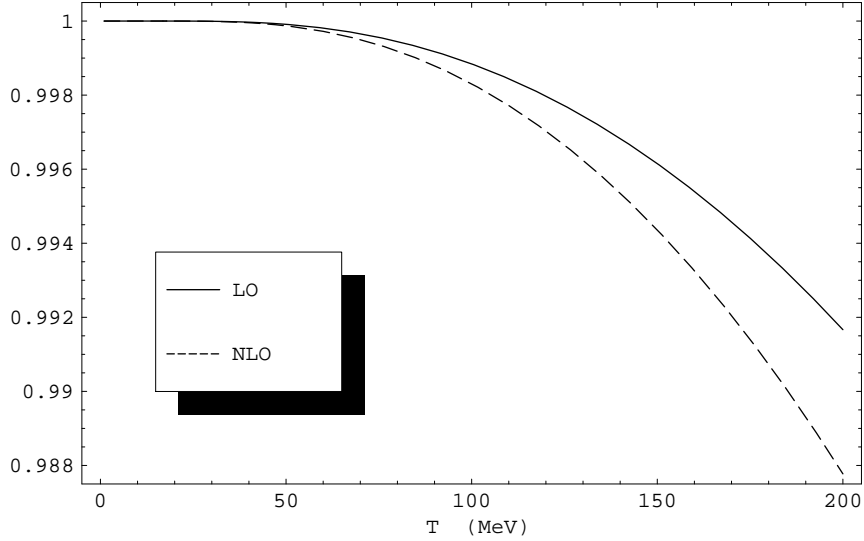,height=8cm}}
\caption{Leading order (solid line) and next-to-leading order (dashed line)
contributions to $M_\eta(T)$. Both curves are normalized to $M_\eta(0)$.}
\label{fig2}
\end{figure}
with $I_2(T) = - M_\pi^2 I_1(T)$. As usual, we drop the ultraviolet divergent
part of the pion loops as these can in principle be reabsorbed in vacuum parameters, 
including next-to-next-to-leading order counter-terms. 
The sum over $n$ can then be readily evaluated using standard 
techniques~\cite{Kapusta},
\begin{equation}
\label{ione}
I_1(T) = \int {d^3\vec k \over (2\pi)^3}\, {1 \over \omega}\, 
{1\over \exp(\omega/T) -1} = 
{M_\pi T \over 2 \pi^2} \sum_{n=0}^\infty {1 \over n} 
K_1 \Big({n M_\pi \over T}\Big)\ ,
\end{equation}
where $\omega^2 = \vec k^2 + M_\pi^2$.
For instance, for $T \gsim M_\pi$
\begin{equation}
I_1(T) \approx {T^2\over 12}\ .
\end{equation}
In the sequel we  simply compute (\ref{ione}) numerically.
As the relevant formula are not particularly transparent, we have relegated them to 
the appendix. Fig.~\ref{fig2} and Fig.~\ref{fig3} show the shift of $M_\eta^\prime$ 
and $M_\eta$ at low temperature both at leading and next-to-leading order. 
To be definite we have chosen the set of parameters corresponding to 
$\theta \simeq -20^\circ$.  
The net thermal effects are not dramatic: both masses decrease, but only slightly. 
As expected, the shift of the mass $M_\eta^\prime$ is more pronounced at 
next-to-leading order, but the effect is not very significant. Again, this is because, 
both at leading and next-to-leading orders, the thermal corrections are chirally 
suppressed, $\propto M_\pi^2 T^2/f_\pi^2$. 
For completeness, we have also plotted in Fig.~\ref{fig4} the shift of the mixing 
angle at low temperature. As both the $\eta$ and $\eta^\prime$ masses diminish, the 
angle is not very much affected. It decreases a bit (toward ideal mixing?), consistent 
with the relatively larger shift of $M_{\eta^\prime}$.
\begin{figure}[t]
\centerline{\epsfig{file=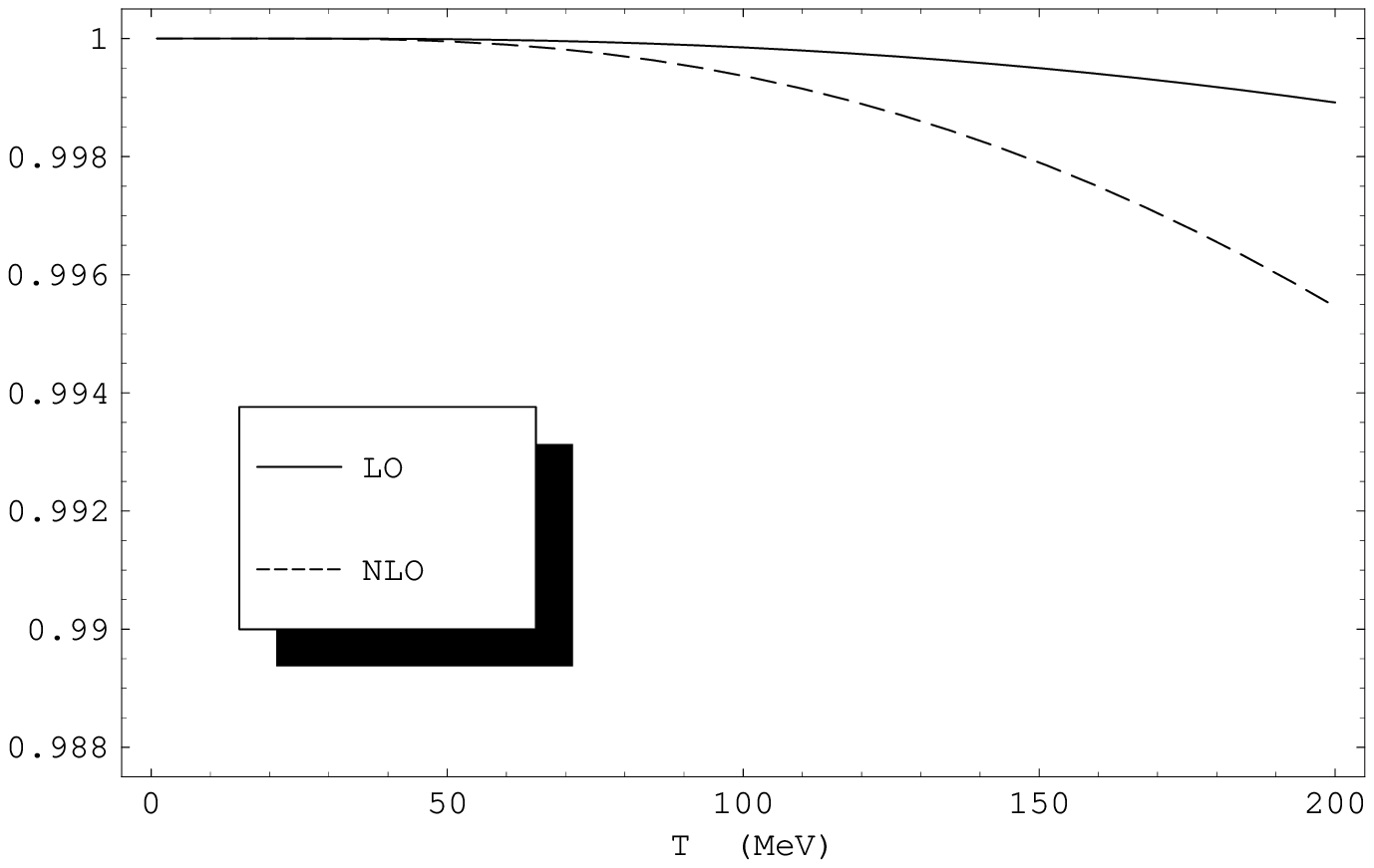,height=8cm}}
\caption{Leading order (solid line) and next-to-leading order (dashed line)
contributions to $M_{\eta^\prime}(T)$. Both curves are normalized to 
$M_{\eta^\prime}(0)$.}
\label{fig3}
\end{figure}

\section{Lessons from large $N_c$?}

\begin{figure}[t] 
\centerline{\epsfig{file=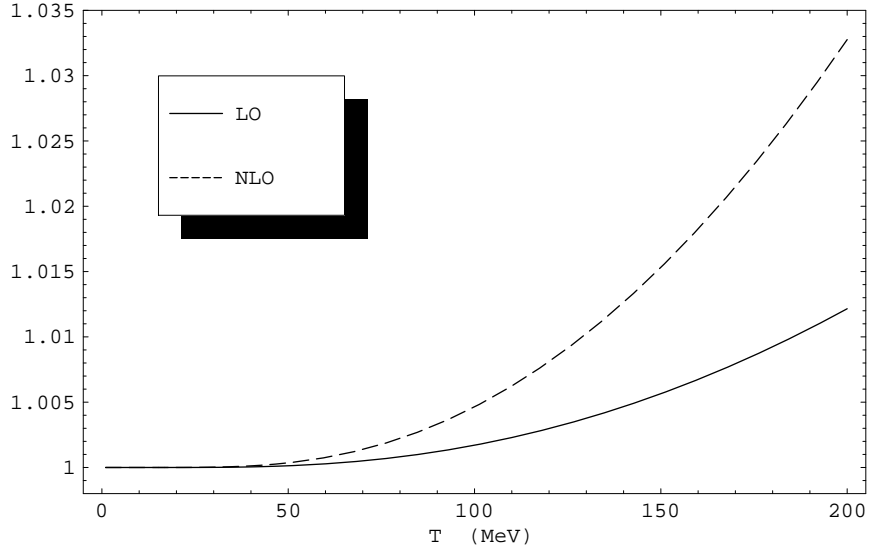,height=8cm}}
\caption{Leading order (solid line) and next-to-leading order (dashed line)
contributions to $\tan 2\theta(T)$. Both curves are normalized to 
$\tan 2\theta(0)$.}
\label{fig4}
\end{figure}
As we have seen in the previous section, the mass of $\eta^\prime$ is almost not 
affected at low temperatures in a pion bath. This is because, at this order, the 
pion thermal corrections are chirally suppressed, smaller than 
$M^2_\pi \simeq 0.02$ GeV$^2$, and thus essentially negligible compared to the 
contribution from the $U(1)_A$ symmetry breaking term $v_{02} \simeq 0.22$ GeV$^2$.
In particular, in the chiral limit, $M_\pi =0$, all the 
corrections vanish and  $M_{\eta^\prime}$ is temperature independent up to 
next-to-leading order in $\chi PT$.
In the chiral limit, the leading contribution from pions to the shift of 
$M_{\eta^\prime}$ presumably arise from ${\cal O}(p^4)$ operators 
like\footnote{The $O(p^2)$ operator 
\begin{equation}
{\cal L} \sim f^2\,{X^2 \over N_c^2}\,
\langle \partial_\mu U^\dagger \partial_\mu U\rangle\ ,
\end{equation}
can  contribute at 
one-loop if and only if $M_\pi^2 \not= 0$. It could contribute at two-loop order 
in the chiral limit $\propto T^4$, but doesn't because pion interactions are too 
soft. This is a well-known feature of pion thermal corrections which is for 
instance manifest in the absence of $T^6$ terms in the free energy of a pion 
gas in the chiral limit \cite{Gerber:1989tt}, or in the fact that massless thermal 
pions move at the speed of light to order $T^2$ \cite{Pisarski:1996ty}.}
\begin{equation}
\label{nnlo}
{\cal L} \sim {1\over N_c} \, X^2\, \langle \partial_\mu U \partial_\nu U^\dagger 
\partial_\mu U \partial_\nu U^\dagger \rangle\ .
\end{equation}
The coupling  is ${\cal O}(1/N_c)$ because there is a factor of $1/N_c^2$ coming 
with $X^2$ and one of $N_c$ from the coupling of the ${\cal O}(p^4)$ operator. 
The best way to see this is to replace the coupling $M_0 \sim N_c$ (or $M_3$) 
from the next-to-leading order lagrangian  by a function of $X$, 
$M_0 \rightarrow M_0(X)$ and expand to second order in $X$, which brings down a
factor of $1/N_c^2$. Because there are at least four pions in the expansion of the 
operator~(\ref{nnlo}), the leading pion thermal correction to $M_{\eta^\prime}$ in 
the chiral limit is a two-loop effect,
\begin{equation}
\label{nnloshift}
\delta M^2_{\eta^\prime}(T) \sim {1\over N_c} \, {T^8\over f^6} 
\sim {1 \over N_c^4} \, {T^8 \over \mu_{\rm hadr}^6}\ .
\end{equation}
We have made the large $N_c$ dependence of the pion decay constant $f^2$ manifest 
by defining $f^2 \sim N_c \mu_{\rm hadr}$. Of course, the sign of the correction is
not known and $M_{\eta^\prime}$ could go up or down. 
Also, if we compare with $M_{\eta^\prime}^2(0) \sim \mu_{\rm hadr}^2/N_c$, 
we infer that $M_{\eta^\prime}$ is quasi-constant for temperatures 
\begin{equation}
\label{Tstar}
T \ll T_\ast \sim N_c^{3/8} \mu_{\rm hadr}\ .
\end{equation}

In the large $N_c$ framework, the natural scale for deconfinement is 
$T_c \sim \mu_{\rm hadr}$, which is also the temperature at which the pions from the 
hadronic gas overlap. It is natural to assume that chiral symmetry restoration takes 
place at the same temperature, driven by the release of ${\cal O}(N_c^2)$ gluon 
degrees of freedom \cite{Smilga:1996cm,Pisarski:1984db}. 
The estimate in Eq.~(\ref{Tstar}) then seems to imply that $M_{\eta^\prime}$ is 
essentially constant up to the temperature of deconfinement, since 
$T_\ast \gg T_c \sim \mu_{\rm hadr}$ for $N_c$ large. 
This conclusion is however premature because the low momentum expansion breaks 
down near $T_c$ and we must take into account the contribution of operators with 
arbitrary number of derivatives.
We claim that the dominant operators at large $N_c$ are of the form
\begin{equation} 
\label{leadop}
{\cal L} \sim {1 \over N_c \, \mu_{\rm hadr}^{2 k-4}} X^2 \, 
\langle \partial_{\mu_1} U \partial_{\mu_2} U^\dagger \ldots 
\partial_{\mu_k} U \partial_{\mu_1} U^\dagger \partial_{\mu_2} U \ldots 
\partial_{\mu_k} U^\dagger \rangle\ .
\end{equation}
These operators are irrelevant at low energies but become marginal for 
$\partial \sim T_c$.
A six-derivative operator, for instance, first contribute at three loops 
$\delta M_{\eta^\prime}^2 \propto T^{12}/(N_c^5\mu_{\rm hadr}^{10})$. 
For comparison, the  contribution of a three-loop diagram with a four-derivatives 
(NLO) and a two-derivatives vertices (LO) is 
$\propto 1/N_c \, T^{10}/f^8 \sim 1/N_c^5 \, T^{10}/\mu_{\rm hadr}^8$ and is 
subdominant at large $N_c$. 
For generic $k$, the operators of Eq.~(\ref{leadop}) give 
$\delta M_{\eta^\prime}^{2} \sim T^{4 k}/(N_c^{k +2} \mu_{\rm hadr}^{4k -2})$. 
The ratio of two consecutive terms $k$ and $k+1$ becomes ${\cal O}(1)$ at 
$T^\prime \sim N_c^{1/4} \mu_{\rm hadr}$, independent of $k$. At large $N_c$, 
$T_c \ll T^\prime \ll T^\ast$ and  the perturbative expansion still breaks down 
above the temperature of deconfinement. 

Another set of operators could  be relevant at large $N_c$ because the 
$\eta^\prime$ is then rather light, 
$M_{\eta^\prime}\sim \mu_{\rm hadr}/N_c^{1/2} < T_c$.
Thus one should include operators that involve arbitrary powers of the 
$\eta^\prime$ field, like
\begin{equation}
\begin{array}{rcl}
{\cal L} &\sim& f^2 F\left({X\over N_c}\right) \, 
\langle \partial_\mu U^\dagger \partial_\mu U\rangle \\[1ex]
&=& \left({\eta_0^2\over \mu^2 N_c^2} + {\eta_0^4 \over \mu^4 N_c^5} + 
\ldots \right) \, (\partial_\mu \eta^0)^2 + \mbox{\rm pion terms}\ ,
\end{array}
\end{equation}
which contributes to the wave-function renormalization of $\eta^\prime$,
\begin{equation}
\delta Z_{\eta^\prime} \sim T^2/(N_c^2 \mu_{\rm hadr}^2) + 
T^4/(N_c^5 \mu_{\rm hadr}^4) +\ldots\ ,
\end{equation}
or terms of the form
\begin{equation}
{\cal L} \sim N_c^2 \mu_{\rm hadr}^4 G\left({X\over N_c}\right) \sim 
{\mu_{\rm hadr}^2\over N_c} \eta_0^2 +  \#_1 {1\over N_c^4}\, \eta_0^4 + 
\#_2 {1\over \mu^2_{\rm hadr} N_c^7}\, \eta^6_0 + \ldots\ .
\end{equation}
However, a common feature of these operators is that they are very suppressed at
large $N_c$. They become important only for $T \sim N_c^{3/2} \mu_{\rm hadr}$, 
much higher than $T^\prime$ so that their contribution is subleading compared to 
operators like in Eq.~(\ref{leadop}). 

Can we conclude anything from these considerations?
In all the cases discussed above, the leading thermal corrections to 
$M_{\eta^\prime}$, in the chiral limit and for $N_c$ large, become important for 
temperature which are higher than the critical temperature of deconfinement 
$T_c \sim \mu_{\rm hadr}$ by a factor of $N_c^\gamma$. Although the value of 
$\gamma$ is hard to guess, as various corrections can get mixed up, we believe it 
is reasonable to conjecture that $\gamma$ is strictly positive. 
This implies that just below $T_c$, $M_{\eta^\prime}(T) = M_{\eta^\prime}(0)$ to 
a very good approximation. 
The standard lore is that the deconfining phase transition at $T_c$ is of first 
order for $N_c$ large~\cite{Pisarski:1984db}\footnote{Various arguments, including 
recent developments in string theory (see Sect.~6.2.2 in Ref~\cite{Aharony:2000ti})
and lattice simulations  of $N_c=4$ pure Yang-Mills theory~\cite{Ohta:1999wq}, 
favour a first order deconfining phase transition. 
A case for a second order phase transition has been  made in 
Ref.~\cite{Pisarski:1997yh}, in light of the structure of the Columbia  diagram.}.
Because the temperature at which {\em hadronic} interactions can affect 
$M_{\eta^\prime}$ is (very much) larger than the temperature of deconfinement, we 
expect that changes in $M_{\eta^\prime}$ will be instead triggered by the release 
of the large number of gluons and will thus drop {\em discontinuously} at $T_c$, 
{\em i.e.}~that there is a first order transition to a phase with (approximate)
$U(1)_A$ symmetry. 

This behaviour is not inconsistent with various other expectations. For three
light quark flavours, $N_f \geq 3$, the transition to the chirally symmetric phase 
is probably first order while for $N_f=2$, the phase transition is supposed
to be of second order, in the universality class of $O(4)$~\cite{Pisarski:1984ms}. 
It has been argued by Smilga that the latter behaviour is not inconsistent with a 
first order deconfining phase transition at large $N_c$~\cite{Smilga:1996cm}. 
The reason is that, unlike for $M_{\eta^\prime}$, there is an infinite subset of 
thermal corrections that contribute to the {\em same order} in $N_c$ to the shift 
of the quark condensate $\Sigma \equiv \langle \bar q q\rangle$,
\begin{equation}
\label{chiral}
\Sigma (T) = \Sigma \, \left( 1 - \# {T^2 \over N_c^2 \mu_{\rm hadr}^2} 
F\left({T\over \mu_{\rm hadr}}\right) \right)\ .
\end{equation}
Even though thermal corrections are suppressed like $1/N_c^2$, the (unknown) 
function $F(x)$ may be singular near, but below $T_c\sim \mu_{\rm hadr}$. 
If, for instance, $F$ has a simple pole at $T = T_0 < T_c$, 
$F \sim \mu_{\rm hadr}^2/(T^2 - T_0^2)$ and the chiral phase transition is second 
order with a critical region near $T_c$, that is of order
\begin{equation}
{\Delta T\over T_c} \sim {1\over N_c^2}\ .
\end{equation}
Alternatively, if $U(1)_A$ symmetry is effectively restored at $T_c$, the large 
$N_c$ behaviour (\ref{chiral}) is also consistent with a fluctuation induced first 
order phase transition. 
For $N_f =1$ finally, chiral symmetry is broken by the anomaly at all temperature 
and there is no chiral phase transition. 
However, if  instanton transitions are strongly suppressed just above $T_c$, 
chiral symmetry can be effectively restored and the phase transition is presumably 
first order. 
\begin{figure}[t] 
\centerline{\epsfig{file=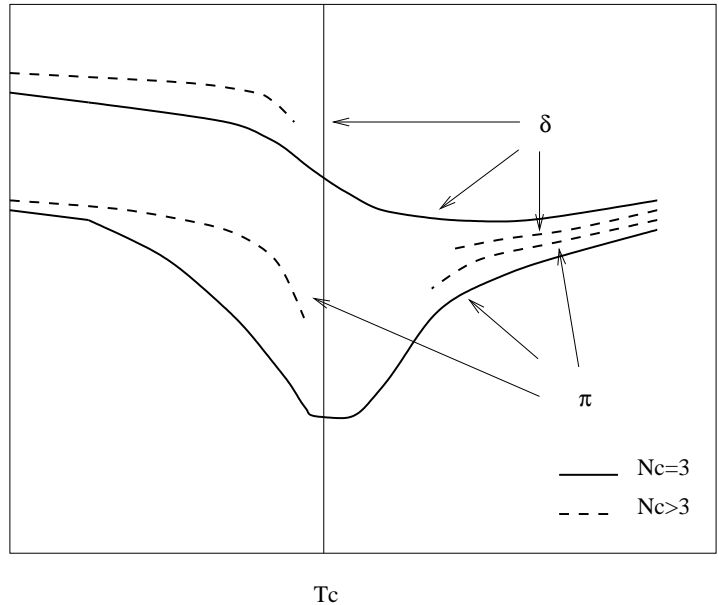,height=8cm}}
\caption{Plot of the $\pi$ and $\delta$ ($a_0$) near the critical
temperature for $N_c=3$ (continuous lines) and possible changes for 
$N_c \gg 3$ (dashed lines).}
\label{fig5}
\end{figure}

\section{Conclusions}

\begin{figure}[t] 
\centerline{\epsfig{file=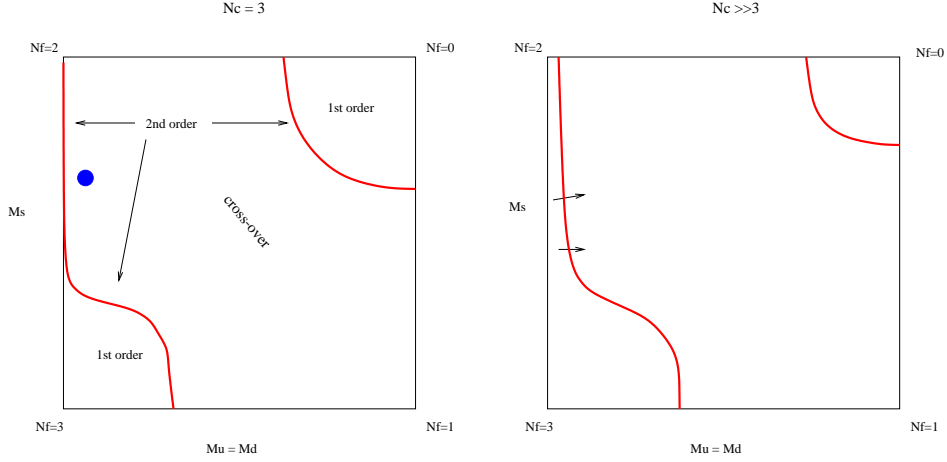,height=6cm}}
\caption{Columbia phase diagram as function of $M_u=M_d$ (horizontal axis) 
and $M_s$ (vertical axes) for $N_c=3$ (left, the blue dot is where QCD 
stands) and how it could evolve at large $N_c$ (right).}
\label{fig6}
\end{figure}

We have studied the behaviour of the mass of the $\eta^\prime$ pseudoscalar meson 
at finite temperature using constraints from chiral symmetry and large $N_c$ power 
counting. The main conclusion to be drawn from this work is that $M_{\eta^\prime}$
is essentially unchanged at low temperatures. 
A tentative analysis of the effect of leading higher order corrections at large 
$N_c$ {\em suggests} that $M_{\eta^\prime}$ changes discontinuously at the 
temperature of deconfinement. The implications of these considerations for the real 
world, {\em i.e.}~$N_c=3$, are not quite clear as we would expect the suppression 
of instanton effects only at asymptotically high temperatures.
It is however striking that recent lattice simulations, with $N_f=2$ 
staggered~\cite{Karsch:1999vy} and domain wall~\cite{vranas} fermions, both show a 
strong suppression of $U(1)_A$ breaking effects at low temperatures 
$T \sim 1.2 \, T_c$. 
Because this temperature is outside the critical region, the order of the chiral 
phase transition is probably not affected.
It could  be of interest to consider doing simulations with $N_c > 3$, although 
this would probably be time consuming, or maybe with one flavour and various $N_c$.
Consider for example a plot of the $\pi$ and $\delta$ susceptibilities near the 
critical temperature as computed on the lattice~\cite{Karsch:1999vy}. 
Large $N_c$ arguments suggest that the curves of the susceptibilities would be 
flatter below $T_c$ ---because the confined phase is colder--- and that  
splitting between $\pi$ and $\delta$ ({\em a.k.a.}~$a_0$) should be narrower above 
$T_c$ ---because $U(1)_A$ breaking is more suppressed, maybe like in 
Fig.~\ref{fig5}---. The Columbia diagram could change accordingly: 
the critical line around the region of small $M_u=M_d$ masses would move as
in Fig.~\ref{fig6}.

\section*{Acknowledgements}

We would like to thank P.~Herrera-Sikl\'ody and F.~Karsch for useful discussions
and S.~Peris and J.~Taron for a careful reading of the manuscript.
Work partly supported by the EEC, TMR-CT98-0169, EURODAPHNE network.

\section*{Appendix}

\subsection{Useful formul\ae}

\subsubsection{Leading order}

\begin{eqnarray}
2mB &=& M_\pi^2\ ,\\[1ex]
x   &=& 2\frac{M_K^2}{M_\pi^2}-2\ ,\\[1ex]
f   &=& f_\pi\ ,\\[1ex]
-3v_{02} &=& M_{\eta\prime}^2-\frac{2M_K^2+M_{\pi}^2}{3}
+\frac{2\sqrt{2}}{3}(M_K^2-M_{\pi}^2)\tan\theta\ .
\end{eqnarray}

\subsubsection{Next-to-leading order}

\paragraph{Some definitions:}

$\Delta_M$, $\Delta_N$ are defined as

\begin{eqnarray}
\Delta_M &\equiv& \frac{8}{f^2}(M_K^2-M_{\pi}^2)(2L_8-L_5)\ ,\\[1ex]
\Delta_N &\equiv& 3v_{31}-12\frac{L_5}{f^2}v_{02}\ .
\end{eqnarray}

\paragraph{Next-to-leading order parameters:} 

The next-to-leading order parameters can be expressed in terms of observables 
through

\begin{eqnarray}
2mB &=& M_\pi^2\left(1-\frac{M_{\pi}^2}{M_K^2-M_{\pi}^2}
                     \Delta_M\right)\ ,\\[1ex]
x   &=& 2\frac{M_K^2}{M_\pi^2}(1-\Delta_M)-2\ ,\\[1ex]
f   &=& f_\pi\left(1-4\frac{L_5}{f^2}M_{\pi}^2\right)\ ,\\[1ex]
-3v_{02} &=& M_{\eta\prime}^2-\frac{2M_K^2+M_{\pi}^2}{3}\nonumber\\[1ex]
&+& \frac{2\sqrt{2}}{3}(M_K^2-M_{\pi}^2)(1+\Delta_M-\Delta_N)
    \tan\theta\nonumber\\[1ex]
&-& \frac{2}{3}\left[(M_K^2-M_{\pi}^2)\Delta_M-
                      (2M_K^2+M_{\pi}^2)\Delta_N\right]\ ,\\[1ex]
\frac{L_5}{f^2} &=& \frac{1}{4(M_K^2-M_{\pi}^2)}\Delta_P\ ,
\end{eqnarray}

where

\begin{eqnarray}
\Delta_M &=& \frac{M_{\pi}^2+3M_{\eta}^2-4M_K^2
                   +3(M_{\eta\prime}^2-M_{\eta}^2)\sin^2\theta}
                  {4(M_K^2-M_{\pi}^2)}\ ,\\[1ex]
\Delta_N &=& 1+\frac{3}{4\sqrt{2}}
             \frac{(M_{\eta\prime}^2-M_{\eta}^2)\sin2\theta}
                  {M_K^2-M_{\pi}^2}+\Delta_M\ ,\\[1ex]
\Delta_P &=& \frac{f_K}{f_\pi}-1\ .
\end{eqnarray}

\subsection{Results at $T=0$}

\subsubsection{Leading order}

\paragraph{Mass matrix:} 

\begin{eqnarray}
m_{88}^2 &=& \frac{1}{3}(4M_K^2-M_{\pi}^2)\ ,\\[1ex]
m_{80}^2 &=& -\frac{2\sqrt{2}}{3}(M_K^2-M_{\pi}^2)\ ,\\[1ex]
m_{00}^2 &=& \frac{1}{3}(2M_K^2+M_{\pi}^2)-3v_{02}\nonumber\\[1ex]
         &=& \frac{1}{3}(2M_K^2+M_{\pi}^2)-\frac{2}{3}y(M_K^2-M_{\pi}^2)\ ,
\end{eqnarray}

where

\begin{equation}
y\equiv \frac{9v_{02}}{2(M_K^2-M_{\pi}^2)}\ .
\end{equation}

\paragraph{Mixing angle:} 

\begin{equation}
\tan2\theta\equiv
\frac{2m_{80}^2}{m_{00}^2-m_{88}^2}=\frac{2\sqrt{2}}{1+y}\ .
\end{equation}

\paragraph{Physical masses:} 

\begin{eqnarray}
M_{\eta}^2 &=& M_K^2-\frac{M_K^2-M_{\pi}^2}{3}(y+\sqrt{9+2y+y^2})\ ,\\[1ex]
M_{\eta\prime}^2 &=& M_K^2-\frac{M_K^2-M_{\pi}^2}{3}(y-\sqrt{9+2y+y^2})\ ,
\end{eqnarray}

with

\begin{equation}
M_{\eta}^2+M_{\eta\prime}^2\equiv m_{88}^2+m_{00}^2=2M_K^2-\frac{2}{3}
y(M_K^2-M_{\pi}^2)\ .
\end{equation}

\subsubsection{Next-to-leading order}

\paragraph{Mass matrix:} 

\begin{eqnarray}
m_{88}^2 &=& \frac{1}{3}(4M_K^2-M_{\pi}^2)
             +\frac{4}{3}(M_K^2-M_{\pi}^2)\Delta_M\ ,\\[1ex]
m_{80}^2 &=& -\frac{2\sqrt{2}}{3}(M_K^2-M_{\pi}^2)(1+\Delta_M-\Delta_N)\ ,\\[1ex]
m_{00}^2 &=& \frac{1}{3}(2M_K^2+M_{\pi}^2)-3v_{02}\nonumber\\[1ex]
         &+& \frac{2}{3}(M_K^2-M_{\pi}^2)\Delta_M-
             \frac{2}{3}(2M_K^2+M_{\pi}^2)\Delta_N\nonumber\\[1ex]
         &=& \frac{1}{3}(2M_K^2+M_{\pi}^2)(1-2\Delta_N)+
             \frac{2}{3}(M_K^2-M_{\pi}^2)\Delta_M\nonumber\\[1ex]
         &-& \frac{2}{3}y(M_K^2-M_{\pi}^2)\ .
\end{eqnarray}

\paragraph{Mixing angle:} 

\begin{equation}
\tan2\theta=\frac{2\sqrt{2}}{1+y}\left(1+\frac{y}{1+y}\Delta_M-
\left(1+\frac{1}{1+y}\frac{2M_K^2+M_{\pi}^2}{M_K^2-M_{\pi}^2}\right)\Delta_N\right)\ .
\end{equation}

\paragraph{Physical masses:} 

\begin{eqnarray}
M_{\eta}^2 &=& M_K^2-\frac{M_K^2-M_{\pi}^2}{3}(y+\sqrt{9+2y+y^2})\nonumber\\[1ex]
&+& \left(1-\frac{9+y}{3\sqrt{9+2y+y^2}}\right)(M_K^2-M_{\pi}^2)\Delta_M\nonumber\\[1ex]
&-& \frac{1}{3}\left(2M_K^2+M_{\pi}^2-
\frac{3(2M_K^2-3M_{\pi}^2)-y(2M_K^2+M_{\pi}^2)}{\sqrt{9+2y+y^2}}\right)\Delta_N\ ,\\[1ex]
M_{\eta\prime}^2 &=& M_K^2-\frac{M_K^2-M_{\pi}^2}{3}(y-\sqrt{9+2y+y^2})\nonumber\\[1ex]
&+& \left(1+\frac{9+y}{3\sqrt{9+2y+y^2}}\right)(M_K^2-M_{\pi}^2)\Delta_M\nonumber\\[1ex]
&-& \frac{1}{3}\left(2M_K^2+M_{\pi}^2+
\frac{3(2M_K^2-3M_{\pi}^2)-y(2M_K^2+M_{\pi}^2)}{\sqrt{9+2y+y^2}}\right)\Delta_N\ ,
\end{eqnarray}

with

\begin{eqnarray}
M_{\eta}^2+M_{\eta\prime}^2 &=& 2M_K^2-\frac{2}{3}y(M_K^2-M_{\pi}^2)\nonumber\\[1ex]
&+& 2(M_K^2-M_{\pi}^2)\Delta_M-\frac{2}{3}(2M_K^2+M_{\pi}^2)\Delta_N\ .
\end{eqnarray}

\subsection{Results at $T\neq 0$}

\subsubsection{Leading order}

\paragraph{Mass matrix:} 

\begin{eqnarray}
m_{88}^2(T) &=& m_{88}^2(0)-\frac{M_{\pi}^2}{2f_{\pi}^2}I(T)\ ,\\[1ex]
m_{80}^2(T) &=& m_{80}^2(0)-\frac{M_{\pi}^2}{\sqrt{2}f_{\pi}^2}I(T)\ ,\\[1ex]
m_{00}^2(T) &=& m_{00}^2(0)-\frac{M_{\pi}^2}{f_{\pi}^2}I(T)\ ,
\end{eqnarray}

where

\begin{eqnarray}
I(T) &\equiv& \int\frac{d^3\vec{k}}{(2\pi)^3}\frac{1}{\omega}\frac{1}{e^{\beta\omega}-1}\ ;
\ \omega=\sqrt{\vec{k}^2+M_{\pi}^2},\ \beta=\frac{1}{T}\nonumber\\[1ex]
&=& \frac{M_{\pi}T}{2\pi^2}\sum_{n=1}^\infty \frac{1}{n}K_1\left(\frac{nM_{\pi}}{T}\right)
\stackrel{M_{\pi}\rightarrow 0}{\longrightarrow}\frac{T^2}{12}\ .
\end{eqnarray}

\paragraph{Mixing angle:} 

\begin{equation}
\tan2\theta(T)=\tan2\theta(0)+\frac{2\sqrt{2}}{1+y}
\frac{y}{1+y}\frac{3}{4}\frac{M_{\pi}^2}{M_K^2-M_{\pi}^2}\frac{1}{f_{\pi}^2}I(T)\ .
\end{equation}

\paragraph{Physical masses:}

\begin{eqnarray}
M_{\eta}^2(T) &=& M_{\eta}^2(0)-\frac{3}{4}\frac{M_{\pi}^2}{f_{\pi}^2}I(T)
\left(1+\frac{9+y}{3\sqrt{9+2y+y^2}}\right)\ ,\\[1ex]
M_{\eta\prime}^2(T) &=& M_{\eta\prime}^2(0)-\frac{3}{4}\frac{M_{\pi}^2}{f_{\pi}^2}I(T)
\left(1-\frac{9+y}{3\sqrt{9+2y+y^2}}\right)\ ,
\end{eqnarray}

with

\begin{equation}
M_{\eta}^2(T)+M_{\eta\prime}^2(T)=M_{\eta}^2(0)+M_{\eta\prime}^2(0)
-\frac{3}{2}\frac{M_{\pi}^2}{f_{\pi}^2}I(T)\ .
\end{equation}

\subsubsection{Next-to-leading order}

\paragraph{Mass matrix:} 

\begin{eqnarray}
m_{88}^2(T) &=& m_{88}^2(0)-\frac{M_{\pi}^2}{2f_{\pi}^2}I(T)
\left(1+\frac{2M_{\pi}^2}{M_K^2-M_{\pi}^2}\left(\Delta_P+\frac{3}{2}\Delta_M\right)\right.
\nonumber\\[1ex]
&+& \left.24\frac{M_{\pi}^2}{f_{\pi}^2}(M_0+M_3)\right)\ ,\\[1ex]
m_{80}^2(T) &=& m_{80}^2(0)-\frac{M_{\pi}^2}{\sqrt{2}f_{\pi}^2}I(T)
\left(1+\frac{2M_{\pi}^2}{M_K^2-M_{\pi}^2}\left(\Delta_P+\frac{3}{2}\Delta_M\right)\right.
\nonumber\\[1ex]
&-& \left.\Delta_N+24\frac{M_{\pi}^2}{f_{\pi}^2}(M_0+M_3)
          \left(1-\frac{y}{3}\frac{M_K^2-M_{\pi}^2}{M_{\pi}^2}\right)\right)\ ,\\[1ex]
m_{00}^2(T) &=& m_{00}^2(0)-\frac{M_{\pi}^2}{f_{\pi}^2}I(T)
\left(1+\frac{2M_{\pi}^2}{M_K^2-M_{\pi}^2}\left(\Delta_P+\frac{3}{2}\Delta_M\right)\right.
\nonumber\\[1ex]
&-& \left.2\Delta_N+24\frac{M_{\pi}^2}{f_{\pi}^2}(M_0+M_3)
          \left(1-\frac{2y}{3}\frac{M_K^2-M_{\pi}^2}{M_{\pi}^2}\right)\right)\ .
\end{eqnarray}

\paragraph{Mixing angle:}

\begin{eqnarray}
\tan2\theta(T) &=& \tan2\theta(0)+\frac{2\sqrt{2}}{1+y}
\frac{y}{1+y}\frac{3}{4}\frac{M_{\pi}^2}{M_K^2-M_{\pi}^2}\frac{1}{f_{\pi}^2}I(T)\times
\nonumber\\[1ex]
\left(1\right. &+& \frac{2M_{\pi}^2}{M_K^2-M_{\pi}^2}\left(\Delta_P+\frac{3}{2}
\Delta_M\right)
-\frac{2}{1+y}\Delta_M\nonumber\\[1ex]
&+& \frac{1-y}{1+y}\Delta_N+
\frac{3}{(1+y)y}\frac{2M_K^2-(1+y)M_{\pi}^2}{M_K^2-M_{\pi}^2}\Delta_N\nonumber\\[1ex]
&+& \left.24\frac{M_{\pi}^2}{f_{\pi}^2}(M_0+M_3)
          \left(\frac{M_K^2-\frac{y}{3}(M_K^2-M_{\pi}^2)}{M_{\pi}^2}\right)\right)\ .
\end{eqnarray}

\paragraph{Physical masses:}

\begin{eqnarray}
M_{\eta}^2(T) &=& M_{\eta}^2(0)-\frac{3}{4}\frac{M_{\pi}^2}{f_{\pi}^2}I(T)
\left(1+\frac{9+y}{3\sqrt{9+2y+y^2}}\right.\nonumber\\[1ex]
&+& \frac{2M_{\pi}^2}{M_K^2-M_{\pi}^2}\left(\left(1+\frac{9+y}{3\sqrt{9+2y+y^2}}\right)
\Delta_P\right.\nonumber\\[1ex]
& & +\frac{3}{2}\left(1+\frac{27(3+y)+(9+y)y^2}{3(9+2y+y^2)^{3/2}}\right.\nonumber\\[1ex]
& & \left.\left.+\frac{\frac{2}{3}y^2}{3(9+2y+y^2)^{3/2}}\frac{4M_K^2-M_{\pi}^2}{M_{\pi}^2}
\right)\Delta_M\right)\nonumber\\[1ex]
&-& \frac{4}{3}\left(1+\frac{(3+y)((9+y)+(3+y)y)}{(9+2y+y^2)^{3/2}}\right.\nonumber\\[1ex]
& & \left.+\frac{6y}{(9+2y+y^2)^{3/2}}\frac{M_{\pi}^2}{M_K^2-M_{\pi}^2}\right)\Delta_N
\nonumber\\[1ex]
&+& 24\frac{M_{\pi}^2}{f_{\pi}^2}(M_0+M_3)
    \left(1+\frac{9+y}{3\sqrt{9+2y+y^2}}\right.\nonumber\\[1ex]
& & \left.\left.-\frac{4}{9}y\left(1+\frac{3+y}{\sqrt{9+2y+y^2}}\right)
\frac{M_K^2-M_{\pi}^2}{M_{\pi}^2}\right)\right)\ ,\\[1ex]
M_{\eta\prime}^2(T) &=& M_{\eta\prime}^2(0)-\frac{3}{4}\frac{M_{\pi}^2}{f_{\pi}^2}I(T)
\left(1-\frac{9+y}{3\sqrt{9+2y+y^2}}\right.\nonumber\\[1ex]
&+& \frac{2M_{\pi}^2}{M_K^2-M_{\pi}^2}\left(\left(1-\frac{9+y}{3\sqrt{9+2y+y^2}}\right)
\Delta_P\right.\nonumber\\[1ex]
& & +\frac{3}{2}\left(1-\frac{27(3+y)+(9+y)y^2}{3(9+2y+y^2)^{3/2}}\right.\nonumber\\[1ex]
& & \left.\left.-\frac{\frac{2}{3}y^2}{3(9+2y+y^2)^{3/2}}\frac{4M_K^2-M_{\pi}^2}{M_{\pi}^2}
\right)\Delta_M\right)\nonumber\\[1ex]
&-& \frac{4}{3}\left(1-\frac{(3+y)((9+y)+(3+y)y)}{(9+2y+y^2)^{3/2}}\right.\nonumber\\[1ex]
& & \left.-\frac{6y}{(9+2y+y^2)^{3/2}}\frac{M_{\pi}^2}{M_K^2-M_{\pi}^2}\right)\Delta_N
\nonumber\\[1ex]
&+& 24\frac{M_{\pi}^2}{f_{\pi}^2}(M_0+M_3)
    \left(1-\frac{9+y}{3\sqrt{9+2y+y^2}}\right.\nonumber\\[1ex]
& & \left.\left.-\frac{4}{9}y\left(1-\frac{3+y}{\sqrt{9+2y+y^2}}\right)
\frac{M_K^2-M_{\pi}^2}{M_{\pi}^2}\right)\right)\ ,
\end{eqnarray}

with

\begin{eqnarray}
M_{\eta}^2(T)+M_{\eta\prime}^2(T) &=& M_{\eta}^2(0)+M_{\eta\prime}^2(0)
-\frac{3}{2}\frac{M_{\pi}^2}{f_{\pi}^2}I(T)\times\nonumber\\[1ex]
\left(1\right. &+& \frac{2M_{\pi}^2}{M_K^2-M_{\pi}^2}\left(\Delta_P+\frac{3}{2}\Delta_M\right)
-\frac{4}{3}\Delta_N\nonumber\\[1ex]
&+& \left.24\frac{M_{\pi}^2}{f_{\pi}^2}(M_0+M_3)
          \left(1-\frac{4}{9}y\frac{M_K^2-M_{\pi}^2}{M_{\pi}^2}\right)\right)\ .
\end{eqnarray}

\end{document}